\documentstyle[epsf,rotate]{article}
\def\be{\begin{equation}}					 
\def\ee{\end{equation}}
\def\ber{\begin{eqnarray}}
\def\eer{\end{eqnarray}}	
\def\dint{\mathop{\intop\kern-0.5em\intop}}

\begin{document}

\begin{center}
{\Large \bf Three-body Forces, Single Diffraction Dissociation and
Shadow Corrections to Hadron-Deuteron Total
Cross-Sections}\footnote{The talk presented at the XV 
International Workshop on High Energy Physics and Quantum Field
Theory, Tver, Russia, September 14--20, 2000.} \\

\vspace{4mm}

A.A. Arkhipov\\
{\it Institute for High Energy Physics \\
 142284 Protvino, Moscow Region, Russia}\\
\end{center}

\begin{abstract}
The relationships between the fundamental dynamics and diffraction
phenomena in scattering from two-body composite system are discussed.
A new simple formula for the shadow corrections to the total
cross-section in scattering from deuteron has been derived and new
scaling characteristics with a clear physical interpretation have
been established. The effect of weakening the inelastic screening
at super-high energies is theoretically discovered. A comparison of
the obtained structure for the shadow corrections with the
experimental data on proton(antiproton)-deuteron total cross sections
has been performed. It is shown that there is quite a remarkable
correspondens of the theory with the experimental data. 
\end{abstract}
		   
\section*{Introduction}

From childhood we see a mysterious play of light and shadow which is
really a manifestation of diffraction phenomena. It turns out that
diffraction phenomena take place in the processes with particles and
nuclei as well. At present time it is well established  that the
fundamental dynamics of particles and nuclei contains the dynamics of
diffraction phenomena as a special case. Everybody who works in
diffraction (high-energy) physics has learned that profiles of
shadows are related to the fundamental dynamics. So, our intuition
suggests that there are very deep relationships between three blocks
shown on the diagram below
%\vspace{5mm}
\begin{center}
$\framebox{\bf DYNAMICS}$
\end{center}
%\vspace*{1cm}
\begin{center}
$\swarrow \hskip 0.8 true in \nwarrow$
\end{center}
%\vspace*{1cm}
\begin{center}
$\framebox{\bf DIFFRACTION}$
\hskip 0.2 true in
$\Longrightarrow$
\hskip 0.2 true in
$\framebox{\bf SHADOW}$
\end{center}
%\vspace{5mm}
Here I'd like to discuss some aspects of these relationships in
the framework of general structures in the relativistic quantum
theory. It will be shown that the diffraction phenomena with a
shadow dynamics in the scattering of a high-energy particle from 
composite systems, like nuclei, will be characterized by the scaling
laws with a quite clear physical meaning. Deuteron is the simplest
composite nuclear system, that's why it may serve as the best
laboratory to study shadow dynamics. I'll also attempt  to
demonstrate new structures of the shadow dynamics in the light of 
existing experimental data on proton(antiproton)-deuteron total
cross sections. Therefore, above all, let me remind you some
well-known facts to restore what was many years ago. 

First of all, experimental and theoretical studies of high-energy
particle interaction with deuterons have shown that the total cross
section in scattering from deuteron cannot be treated as equal to the
sum of total cross sections in scattering from free proton and
neutron even in the range of asymptotically high energies. Glauber
was the first to propose the explanation of this effect. Using the
methods of diffraction theory, the quasiclassical picture for
scattering from composite systems and eikonal approximation for
high-energy scattering amplitudes, he found long ago  \cite{1} that
the total cross section in scattering  from deuteron could be
expressed by the formula
\be
\sigma_d = \sigma_p +\sigma_n - \delta\sigma, \label{1}
\ee
where
\be
\delta\sigma = \delta\sigma_G =
\frac{\sigma_p\cdot \sigma_n}{4\pi}<\frac{1}{r^2}>_d.\label{2}
\ee
Here $\sigma_d, \sigma_p, \sigma_n$ are the total cross sections
in scattering from deuteron, proton and neutron, $<r^{-2}>_d$ is the
average value for the inverse square of the distance between the
nucleons inside a deuteron, $\delta \sigma_G$ is the Glauber
shadow  correction describing the effect of eclipsing or the
screening effect in the recent terminology. The Glauber shadow
correction  has quite a clear physical interpretation. This
correction originates from elastic rescattering of an incident
particle on the nucleons in a deuteron and corresponds to the
configuration when the relative position of the nucleons in a
deuteron is such that one casts its ``shadow" on the other \cite{1}.

It was soon understood that in the range of high energies the shadow
effects may arise due to inelastic interactions of an incident
particle with the nucleons of a deuteron \cite{2,3,4,5,6}.
Therefore, an inelastic shadow correction had to be added to the
Glauber one.

A simple formula for the total (elastic plus inelastic) shadow
correction had been derived by Gribov \cite{4} in the assumption of
Pomeron dominance in the dynamics of elastic and inelastic
interactions. However, it was observed that the calculations
performed by the Gribov formula did not meet the experimental data:
The calculated values of the inelastic shadow correction
over-estimated the experimental values. 

The idea, that the Pomeron dominance is not justified at the
recently available energies, has been explored in papers \cite{5}.
The authors of Refs. \cite{5} argued that the account of the
triple-reggeon diagrams for six-point amplitude in addition to the
triple-pomeron ones allowed them to obtain a good agreement with the
experiment. Alberi and Baldracchini replied \cite{6} and pointed out
that discrepancy between theory and experiment could not be
eliminated by taking into account the triple-reggeon diagrams: In
fact, it is needed to modify  the dynamics of the six-point amplitude
with more complicated diagrams than the triple Regge ones. This means
that up to now we had not, in the framework of Regge phenomenology, a
clear understanding for the shadow corrections in elastic scattering
from deuteron.

The theoretical understanding of the screening effects in scattering
from any composite system is of fundamental importance, because the
structure of shadow corrections is deeply related to the structure of
the composite system itself. At the same time the structure of the
shadow corrections displays new aspects for the fundamental dynamics.

Here we are concerned with the study of shadow dynamics in scattering
from deuteron in some details. A new simple formula for the shadow
corrections to the total cross-section in scattering from deuteron
will be presented and new  scaling characteristics with a clear
physical interpretation will be established. Furthermore, the effect
of weakening the inelastic screening at super-high energies is
theoretically discovered here. We also made a preliminary
comparison of the obtained structure for the shadow corrections with 
the  experimental data on proton(antiproton)-deuteron total cross
sections. It will be shown that there is quite a remarkable
correspondence of the theory with the experimental data. 

\section{Scattering from deuteron}

In our papers \cite{7,8,9} the problem of scattering from two-body
bound states was treated with the help of dynamic equations obtained
on the basis of single-time formalism in QFT \cite{7}. Now I shall
briefly sketch the basic results of our analysis of high-energy
particle scattering from deuteron. As has been shown in \cite{8,9},
the total cross-section in the scattering from deuteron can be
expressed by the formula
\be
\sigma_{hd}^{tot}(s) = \sigma_{hp}^{tot}(\hat s)
+\sigma_{hn}^{tot}(\hat s) -
\delta\sigma(s), \label{3}
\ee
where $\sigma_{hd}, \sigma_{hp}, \sigma_{hn}$  are the total
cross-sections in scattering from deuteron, proton and neutron, 
\begin{equation}
\delta\sigma(s) = \delta\sigma^{el}(s)
+\delta\sigma^{inel}(s),\label{4}
\end{equation}
\begin{equation}
\delta\sigma^{el}(s) =
\frac{\sigma_{hp}^{tot}(\hat s) \sigma_{hn}^{tot}(\hat s)}{4\pi(
R^2_d+B_{hp}(\hat s)+B_{hn}(\hat s))},\quad \hat s = \frac{s}{2},
\label{5}
\end{equation}
$B_{hN}(s)$ is the slope of the forward diffraction peak in the
elastic scattering from nucleon, $1/R_d^2$ is defined by the deuteron
relativistic formfactor \cite{8}
\be
\frac{1}{R_d^2} \equiv \frac{q}{\pi }\int\frac{d\vec \Delta \Phi
(\vec \Delta )}{2\omega_h (\vec 
q+\vec \Delta )}\delta \left[\omega_h (\vec q+\vec \Delta )-\omega_h
(\vec q\,)\right],\ \ \frac{s}{2M_d}\cong q\cong \frac{\hat
s}{2M_N}, \label{6}
\ee
$\delta\sigma^{el}$ is the shadow correction describing the effect of
eclipsing or screening effect during the elastic rescatterings of an
incident hadron on the nucleons in a deuteron.

The quantity $\delta\sigma^{inel}$ in our approach represents the
contribution of the three-body forces to the total cross-section in
the scattering from deuteron. For the definition of three-body forces
in relativistic quantum theory see recent paper \cite{10} and
references therein. For this quantity paper \cite{9} provides  the
following expression:
\be
\delta\sigma^{inel}(s) = -\frac{(2\pi)^3}{q}
\int \frac{d\vec\Delta\Phi(\vec\Delta)}
{2E_p(\vec\Delta /2)2E_n(\vec\Delta /2)}
Im\,R\bigl(s;-\frac{\vec\Delta}{2},\frac{\vec\Delta}{2},\vec q; 
\frac{\vec\Delta}{2},-\frac{\vec\Delta}{2},\vec q \bigr),\label{7}
\ee
where $q$ is the incident particle momentum in the lab system (rest
frame of deuteron), $\Phi(\vec\Delta)$ is the deuteron relativistic
formfactor, normalized to unity at zero, 
\[
E_N(\vec\Delta)=\sqrt{\vec\Delta^2 + M^2_N}\quad N = p, n,
\]
$M_N$ is the nucleon mass. The function $R$ is expressed via the
amplitude of the three-body forces $T_0$ and the amplitudes of
elastic scattering from the nucleons $T_{hN}$ by the relation
\begin{equation}
R = T_0 + \sum_{N=p,n}(T_0G_0T_{hN} + T_{hN}G_0T_0).\label{8}
\end{equation}
A physical reason for the appearance of $\delta\sigma^{inel}$  is
directly connected with the inelastic  interactions of an incident
particle with the nucleons of deuteron. It can be shown that the
contribution of three-body forces to the scattering amplitude from
deuteron is related to the processes of multiparticle production of
inclusive type in the inelastic interactions of the incident particle
with the nucleons of deuteron \cite{8}. This can be done with the
help of the unitarity equation.

To understand the quantity $\delta\sigma^{inel}$ more clearly we may
consider an elementary model for three-body forces. For simplicity,
let us consider the model proposed in \cite{9} where the imaginary
part of the three-body forces scattering amplitude has the form 
\be
Im\,{\cal F}_0(s; \vec p_1, \vec p_2, \vec p_3; \vec q_1, \vec q_2,
\vec q_3)
= f_0(s)
\exp \Biggl\{-\frac{R^2_0(s)}{4} \sum^{3}_{i=1} (\vec p_i-\vec
q_i)^2\Biggr\},\label{9}
\ee
where $f_0(s)$, $R_0(s)$ are free parameters which, in general, may
depend on the total energy of three-body system. This model
assumption is not so significant for our main conclusions but allows
one to make some calculations in a closed form. Indeed, calculating
all the integrals, we obtain for the quantity $\delta\sigma^{inel}$
\cite{9}
\be
\delta\sigma^{inel}(s) =
\frac{(2\pi)^{9/2}f_0(s)\bar\chi(s)}{sM_N[R^2_d+R^2_0(s)]^{3/2}},
\label{10}
\ee 
where
\be
\bar\chi(s) = \frac{\sigma_{hN}(s/2)}{2\pi[B_{hN}(s/2)+
{\bar R}^2_0(s)]}-1,\label{11}
\ee
\be
{\bar R}^2_0(s)=R^2_0(s)(1-\beta), \qquad \beta =
\frac{R^2_0(s)}{4[R^2_0(s)+R^2_d]},\label{12}
\ee
and we suppose that asymptotically
\[
B_{hp}=B_{hn}\equiv B_{hN},\quad \sigma_{hp}^{tot}=\sigma_{hn}^{tot}
\equiv \sigma_{hN}^{tot}.
\]

\section{Three-body forces and single diffraction dissociation}

\begin{figure}[htb]
\begin{center}
\begin{picture}(275,315)
\put(-40,-50){\epsfxsize=12cm \epsfbox{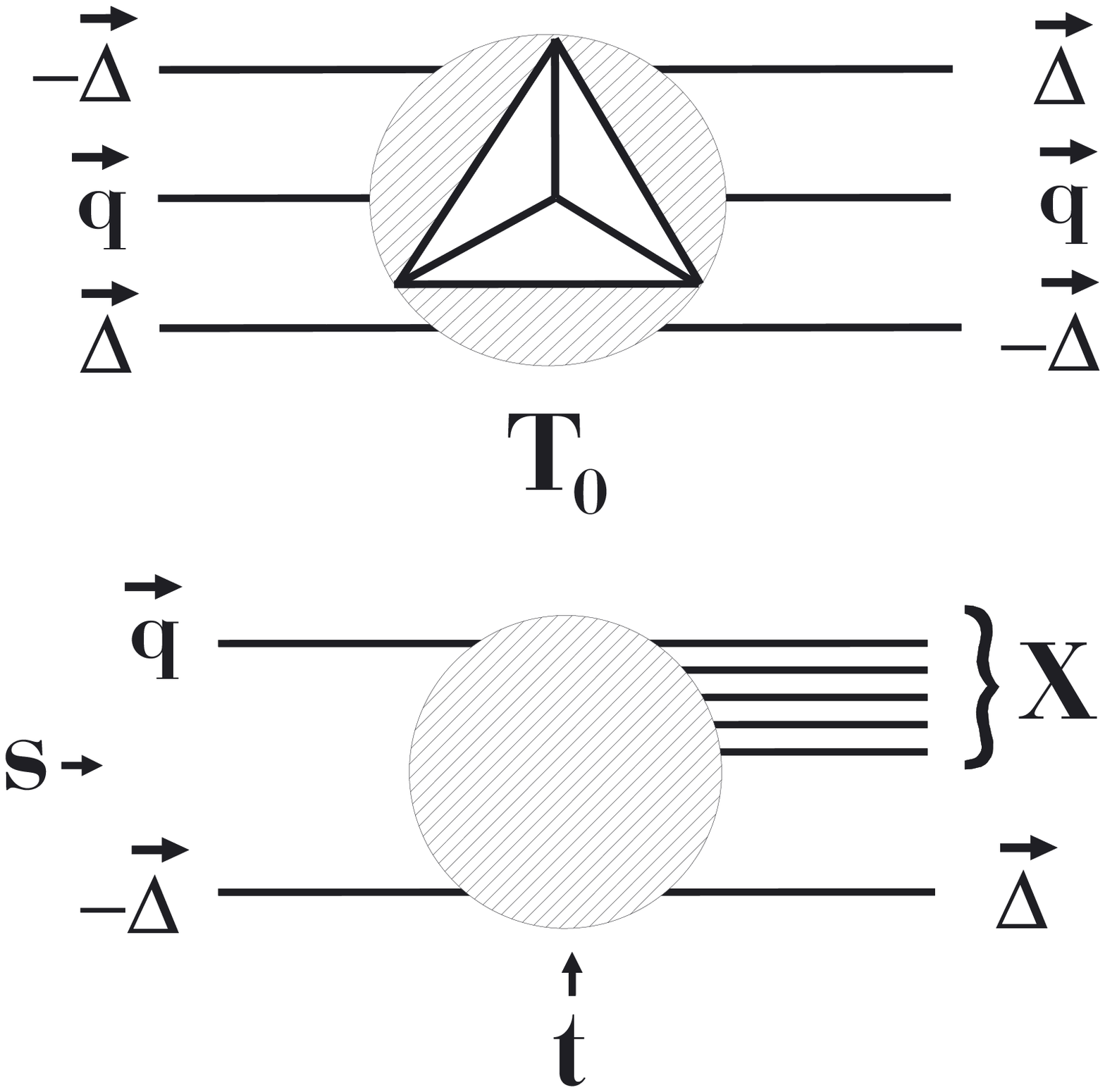}}
\end{picture}
\end{center}

\vspace{2cm}
\caption{Kinematical notations and configuration of
momenta in the relation of one-particle inclusive cross-section to
the three-body forces scattering amplitude.}\label{fig1}
\end{figure}

From the analysis of the problem of high-energy particle
scattering from deuteron we have derived the formula relating
one-particle inclusive cross-section with the imaginary part of the
three-body forces scattering amplitude. This formula looks like
\cite{9,10}
\be 
\fbox{$\displaystyle 2E_N(\vec{\Delta})\frac{{d\sigma}_{hN\rightarrow
NX}}{d\vec{\Delta}}(s,\vec{\Delta}) = 
- \frac{(2\pi)^3}{I(s)}
Im{\cal F}_0^{scr}(\bar s;-\vec{\Delta}, \vec{\Delta}, \vec q; 
\vec{\Delta}, -\vec{\Delta}, \vec q\,)$}\,, \label{13}
\ee 
\vspace{3mm}
\[
Im{\cal F}_0^{scr}(\bar s;-\vec{\Delta}, \vec{\Delta}, \vec q; 
\vec{\Delta}, -\vec{\Delta}, \vec q\,) = Im{\cal F}_0(\bar
s;-\vec{\Delta}, \vec{\Delta}, \vec q; 
\vec{\Delta}, -\vec{\Delta}, \vec q\,)-
\]
\[
-
4\pi\int d\vec{\Delta}'\frac{\delta\left[E_N(\vec{\Delta} -
\vec{\Delta}') + \omega_h(\vec q+\vec{\Delta}') - E_N(\vec{\Delta}) -
\omega_h(\vec q)\right]}{2\omega_{h}(\vec q + \vec{\Delta}')2E_N
(\vec{\Delta} - \vec{\Delta}')}\times
\]
\be
Im{\cal F}_{hN}(\hat s; \vec{\Delta}, \vec q; 
\vec{\Delta}-\vec{\Delta}', \vec q + \vec{\Delta}'\,)Im{\cal
F}_0(\bar s;-\vec{\Delta}, \vec{\Delta}-\vec{\Delta}', \vec q +
\vec{\Delta}'; 
\vec{\Delta}, -\vec{\Delta}, \vec q\,), \label{14}
\ee 
\vspace{3mm}
\[
E_N(\vec{\Delta})=\sqrt{{\vec{\Delta}}^2+M_N^2},\quad
\omega_h(\vec q)=\sqrt{{\vec q}\,^2+m_h^2},\quad
I(s) = 2{\lambda}^{1/2}(s,m_h^2,M_N^2),
\]
\be
\hat s = \frac{\bar s +
m_h^2 - 2M_N^2}{2},\quad
\bar s = 2(s + M_N^2) - M_X^2,\quad t = - 4{\vec\Delta}^2. \label{15}
\ee
I'd like to draw attention to the minus sign in the R.H.S. of
Eq.~(\ref{13}). The simple model for the three-body forces considered
above (see Eq.~(\ref{9})) gives the following result for the
one-particle inclusive cross-section in the region of diffraction
dissociation
\be
\frac{s}{\pi}\frac{d\sigma_{hN\rightarrow NX}}{dtdM_X^2} =
\frac{(2\pi)^3}{I(s)}\chi(\bar s)Im{\cal F}_0(\bar
s;-\vec{\Delta}, \vec{\Delta}, \vec q; 
\vec{\Delta}, -\vec{\Delta}, \vec q\,)
= \frac{(2\pi)^3}{I(s)}\chi(\bar s)f_0(\bar
s)\exp\Biggl[\frac{R_0^2(\bar s)}{2}t\Biggr], \label{16}
\ee 
where
\be
\chi(\bar s) = \frac{\sigma^{tot}_{hN}({\bar s}/2)}{2\pi[B_{hN}({\bar
s}/2) + R_0^2(\bar s)]} -1.\label{17}
\ee
The configuration of particles momenta and kinematical variables are
shown in Fig.~\ref{fig1}. The variable $\bar s$ in the R.H.S. of
Eq.~(\ref{16}) is related to  the kinematical variables of
one-particle inclusive reaction by Eqs.~(\ref{15}).

We may call the quantity $I(s)\chi^{-1}(\bar s)$  a renormalized flux
a l\`a Goulianos. However, it should be pointed out that in our
approach we have a flux of real particles and function $\chi (s)$ has 
quite a clear physical meaning. The function $\chi (s)$ originates
from initial and final states interactions and  describes the
screening effect or the effect of eclipsing of the three-body forces
by two-body ones \cite{9,10}.

If we take the usual parameterization for one-particle inclusive
cross-section in the region of diffraction dissociation
\be
\frac{s}{\pi}\frac{d\sigma}{dtdM_X^2} = A(s.M_X^2)\exp[b(s,M_X^2)t],
\label{18}
\ee 
then we obtain for the quantities $A$ and $b$
\be
A(s,M_X^2) = \frac{(2\pi)^3}{I(s)}\chi(\bar s)f_0(\bar
s),\quad 
b(s,M_X^2) = \frac{R_0^2(\bar s)}{2} \label{19}.
\ee

Eq.~(\ref{19}) shows that the effective radius of three-body forces
is related to the slope of diffraction cone for inclusive diffraction
dissociation processes in the same way as the effective radius of
two-body forces is related to the slope of diffraction cone in
elastic scattering processes. Moreover, it follows from the
expressions
\be
R_0(\bar s) = \frac{r_0}{M_0} \ln \bar s/s'_0,\quad \bar s =
2(s+M_N^2) - M_X^2 \label{20}
\ee
that the slope of diffraction cone for inclusive
diffraction dissociation processes at a fixed energy decreases  with
the growth  of missing mass. This property agrees well qualitatively 
with the experimentally observable picture. Actually, we have even a 
more remarkable fact: Shrinkage or narrowing of diffraction cone for
inclusive diffraction dissociation processes with the growth of
energy at a fixed missing mass and widening of this cone with the
growth of missing mass at a fixed energy is of universal character.
As it follows from Eq.~(\ref{16}) this property is the consequence of
the fact that the one-particle inclusive cross-section depends on the
variables $s$ and $M_X^2$ via one variable $\bar s$ which is a linear
combination of $s$ and $M_X^2$. This peculiar ``scaling" is the 
manifestation of $O(6)$-symmetry of the three-body forces (\ref{9}).
It would be very desirable to experimentally study this new scaling
law related to the symmetry of the new fundamental (three-body)
forces.

Now let us take into account that the functions $\chi$ and $\bar\chi$
are almost the same. In fact, $\chi(s) = \bar\chi(s)$ if the
condition $R_0^2(s) << R_d^2$ is realized, because in that case
$\beta << 1$, but in a general case we have a bound $\beta < 1/4$.
Therefore, we can eliminate one and the same combination $\chi f_0$
entered into equations (\ref{10}), (\ref{16}) and express it through
experimentally measurable quantities. We obtain in this way
\be
A(s,M_X^2) = \frac{\bar s M_N
[R_0^2(\bar
s)+R_d^2]^{3/2}}{(2\pi)^{3/2}I(s)}\delta\sigma^{inel}(\bar s).
\label{21}
\ee
Eq.~(\ref{21}) establishes a deep connection of inelastic shadow
correction with one-particle inclusive cross-section. This relation
allows one to express the inelastic shadow correction via a total
single diffractive dissociation cross-section. This will be done in
the next section.

\section{Elastic and inelastic scaling functions}

We'll start the derivation of the desired expression with the
definition of total single diffractive dissociation cross-section
\be
\sigma_{sd}^{\varepsilon}(s) =
\pi\int_{M_{min}^2}^{\varepsilon
s}\frac{dM_X^2}{s}\int_{t_{-}(M_X^2)}^{t_{+}
(M_X^2)} dt \frac{d\sigma}{dtdM_X^2}. \label{22}
\ee
Here we have specially labeled the total single diffractive
dissociation cross-section by the index $\varepsilon$. It's clear the
parameter $\varepsilon$ defines the range of integration in the
variable $M_X^2$. Unfortunately, there is no common consent
in the choice of this parameter today. However, we would like to
point out an exceptional value for the parameter $\varepsilon$ which
naturally arises from our approach. Namely, let us put
\be
\varepsilon^{ex} = \sqrt{2\pi}/2M_N R_d,\label{ex}
\ee
then we  define the exceptional total single diffractive dissociation
cross-section
\be
\sigma_{sd}^{ex}(s) =
\sigma_{sd}^{\varepsilon}(s)|_{\varepsilon=\varepsilon^{ex}}.
\ee
The exceptional value (\ref{ex}) for the parameter $\varepsilon$ has
a very deep physical meaning: It tells us that the range of
integration in (\ref{22}) in the variable $M_X^2$ is to be determined
by internucleon distances where the two-nucleon bound state may be
organized. The weaker (the larger the internucleon distances)
two-nucleon bound state is, the smaller the range of integration
in (\ref{22}) in the variable $M_X^2$ and vice versa. As a result we
immediately obtain from Eqs.~(\ref{18},
\ref{21},
\ref{22}) \cite{11}
\be
\delta\sigma^{inel}(s) = 2
\sigma_{sd}^{ex}(s)a^{inel}(x_{inel}),\label{23}
\ee
where
\be
a^{inel}(x_{inel}) = \frac{x^2_{inel}}{(1+x^2_{inel})^{3/2}}, \qquad
x^2_{inel} \equiv \frac{R_0^2(s)}{R_d^2} =
\frac{2B_{sd}(s)}{R_d^2},\label{24}
\ee
reminding that $B_{sd}(s) = R_0^2(s)/2 = b(s,M_X^2)|_{M_X^2=2M_N^2}$
\cite{12}.

Here is a convenient place to rewrite the elastic shadow correction
(\ref{5}) in a similar form
\be
\delta\sigma^{el}(s) = 2 \sigma^{el}(s)a^{el}(x_{el}), \qquad
\sigma^{el}(s) \equiv \frac{\sigma_{hN}^{tot\,2}(s)}{16\pi
B_{hN}^{el}(s)}, \label{25}
\ee
where
\be
a^{el}(x_{el}) = \frac{x^2_{el}}{1+x^2_{el}},
\quad
x^2_{el} \equiv \frac{2B_{hN}^{el}(s)}{R_d^2} =
\frac{R_{hN}^2(s)}{R_d^2}, \label{26}
\ee
and we suppose as above that
\[
B_{hp}^{el}=B_{hn}^{el}\equiv B_{hN}^{el},\quad
\sigma_{hp}^{tot}=\sigma_{hn}^{tot}
\equiv \sigma_{hN}^{tot}.
\]
The obtained expressions for the shadow corrections have quite a 
transparent physical meaning, both the elastic $a^{el}$ and
inelastic $a^{inel}$ scaling functions have a clear physical
interpretation. The function $a^{el}$  measures out a portion of
elastic rescattering events among of all the events during the 
interaction of an incident particle with a deuteron as a whole, and
this function attached to the total probability of elastic
interaction of an incident particle with a separate nucleon in a
deuteron. Correspondingly, the function $a^{inel}$ measures out a
portion of inelastic events of inclusive type among of all the events
during  the interaction of an incident particle with a deuteron as a
whole, and this function attached to the total probability of
single diffraction dissociation  of an incident particle on a
separate nucleon in a deuteron. The scaling variables $x_{el}$ and
$x_{inel}$ have quite a clear physical meaning too. The dimensionless
quantity $x_{el}$ characterizes the effective distances measured in
the units of ``fundamental length", which the deuteron size is, in
elastic interactions, but the similar quantity $x_{inel}$
characterizes the effective distances measured in the units of the
same ``fundamental length" during inelastic interactions.

\begin{figure}[htb]
\begin{center}
\begin{picture}(288,178)
\put(20,10){\epsfbox{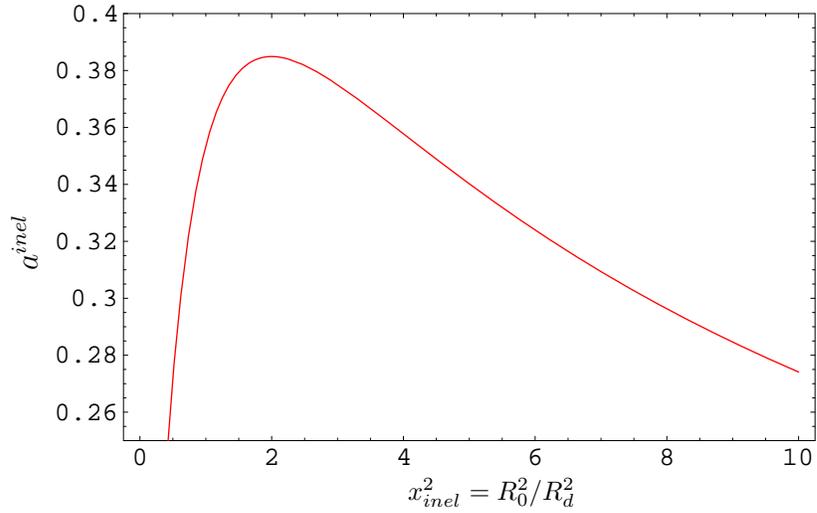}}
\put(155,0){$x^2_{inel}=R_0^2/R_d^2$}
\put(5,87){\rotate{\large$a^{inel}$}}
\end{picture}
\end{center}
\caption{Scaling function $a^{inel}$ versus scaling
variable $x^2_{inel}=R_0^2/R_d^2$.}\label{fig2}
\end{figure}

The functions $a^{el}$ and $a^{inel}$ have quite different
behaviour: $a^{el}$ is a monotonic function while $a^{inel}$ has the 
maximum at the point $x^{max}_{inel}=\sqrt{2}$ where 
$a^{inel}(x^{max}_{inel})=2/3\sqrt{3}$. The graph of $a^{inel}$ is
shown in Fig.~\ref{fig2}. This graph displays an interesting physical
effect of weakening the inelastic eclipsing (screening) at
superhigh energies. The energy at the maximum of $a^{inel}$ can
easily be calculated from the equation $R_0^2(s)=2 R_d^2$ and it will
be done later on.

Account of the real part for the hadron-nucleon elastic scattering
amplitude  modifies the scaling function $a^{el}$ in the following
way:
\be
a^{el}(x_{el}) \longrightarrow
a^{el}(x_{el},\rho_{el})=a^{el}(x_{el})\frac{1-\rho_{el}^2}{1+
\rho_{el}^2},\quad \rho_{el} \equiv \frac{Re{\cal
F}_{hN}^{el}}{Im{\cal F}_{hN}^{el}}.\label{27}
\ee
We see that nonzero value for $\rho_{el}$ violates the scaling
behaviour of $a^{el}$. However, $\rho_{el}$ has a small value at high
energy and moreover $\rho_{el}\rightarrow 0$ at $s \rightarrow
\infty$, therefore, the violation of the scaling law is small at high
energy and we have the restoring scaling in the limit  $s \rightarrow
\infty$.

The scaling function $a^{inel}$ is not modified because all the 
information on the real parts of the amplitudes is contained in the
function $\chi$, which is eliminated in the derivation of formula
(\ref{21}). However, if we would like to speculate in inessential but
subtle distinction between the functions $\chi$ and $\bar\chi$, then
the function $a^{inel}$ should be modified to the form
\be
a^{inel}(x_{inel}) \longrightarrow
a^{inel}(x_{inel},X,\alpha,\beta,\gamma)=a^{inel}(x_{inel})\cdot
r_{\chi}(X,\alpha,\beta,\gamma),
\ee
where
\be
r_{\chi}(X,\alpha,\beta,\gamma)\equiv\frac{\bar\chi}{\chi} =
\frac{[8\alpha X - 1 - 2\gamma(1 - \beta)](1 + 2\gamma)}{(8\alpha X -
1 - 2\gamma)[1 + 2\gamma(1 - \beta)]},
\ee
\[
X\equiv\frac{\sigma^{el}}{\sigma^{tot}},\quad
\alpha\equiv\frac{1-\rho_{el}\rho_0}{1+\rho^2_{el}},\quad \rho_0
\equiv \frac{Re{\cal F}_{0}}{Im{\cal F}_{0}},\quad
\beta\equiv\frac{x^2_{inel}}{4(1+x^2_{inel})},\quad
\gamma\equiv\frac{R^2_0}{2B^{el}}=\frac{B_{sd}}{B^{el}}.
\]
It can easily be seen that
\be
r_{\chi}(0,\alpha,\beta,\gamma)=r_{\chi}(X,0,\beta,\gamma)=r_{\chi}(X
,\alpha,0,\gamma)=r_{\chi}(X,\alpha,\beta,0)=1.
\ee
Besides, we have
\be
0\leq\beta\leq1/4\quad\Longrightarrow\quad 1\leq r_{\chi}\leq{\bar
r}_{\chi},
\quad  {\bar r}_{\chi}=\frac{(8\alpha X - 1 - 3\gamma/2)(1 +
2\gamma)}{(8\alpha X - 1 - 2\gamma)(1 + 3\gamma/2)}.
\ee
From the Froissart bound it follows $\gamma\leq 2$. So, in the case
that $\rho_{el}=0$ or $\rho_{el}=-\rho_0$, taking into account that
$X\leq 1$, we obtain ${\bar r}_{\chi}\leq 5/3$.

Of course, it would be desirable to compare the obtained new
structure for the shadow corrections in elastic scattering from
deuteron with the existing experimental data on hadron-deuteron total
cross sections. The next section will be consecrated to this
comparison.

\section{Comparison with the experimental data}

Here we have tried to make a preliminary comparison of the new
structure for the shadow corrections in elastic scattering from
deuteron with the existing experimental data on proton-deuteron and
antiproton-deuteron total cross sections. To make this comparison in
a more transparent manner, let us rewrite formula (\ref{3}) for the
hadron-deuteron total cross section in a simplified form
\be
\sigma_{hd}^{tot} = 2 \sigma_{hN}^{tot} -
\delta\sigma, \qquad \delta\sigma = \delta\sigma^{el}
+\delta\sigma^{inel},\label{28}
\ee
\be
\delta\sigma^{el} = 2 \sigma^{el}a^{el} = 
\frac{\sigma_{hN}^{tot\,2}}{4\pi(
R^2_d + 2 B_{hN}^{el})}, \label{29}
\ee
\be
\delta\sigma^{inel} =2 \sigma_{sd}^{ex}a^{inel},\quad
a^{inel} = \frac{x^2_{inel}}{(1+x^2_{inel})^{3/2}}, \quad
x^2_{inel} \equiv \frac{R_0^2}{R_d^2}.\label{30}
\ee
All the quantities entered in formulas (\ref{28} -- \ref{30})
are the functions of the energy per nucleon.

\begin{figure}[htb]
\begin{center}
\begin{picture}(288,192)
\put(20,10){\epsfbox{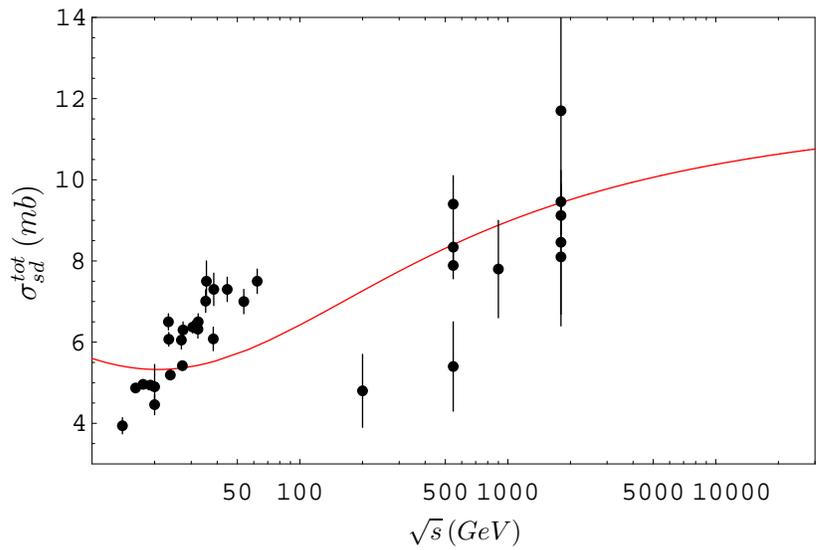}}
\put(155,0){$\sqrt{s}\, (GeV)$}
\put(5,90){\rotate{\large$\sigma^{tot}_{sd}\, (mb)$}}
\end{picture}
\end{center}
\caption{Total single diffraction dissociation cross-section compared 
with the theory (formula (37)). Solid line represents our fit to the
data.}\label{fig3}
\end{figure}

In the first step we analysed the experimental data on
antiproton--deute\-ron total cross sections. We have used  our
theoretical formula describing the global structure of
antiproton-proton total cross sections \cite{10,13} as $\sigma_{\bar
p p}^{tot}=\sigma_{\bar p n}^{tot}\equiv\sigma_{\bar p N}^{tot}$. A 
new fit to the data on the total single diffraction dissociation
cross sections in $\bar p p$ collision with our formula \cite{10}
\be
\sigma_{sd}^{tot}(s) =2 \sigma_{sd}^{ex}(s) = \frac{A_0 +
A_2\ln^2(\sqrt{s}/\sqrt{s_0})}{R_0^2(s)}\label{31}
\ee
has been made as well using a wider set of the data (see Table
\ref{tab}).
The new fit yielded
\[
A_0 = 28.05\pm 0.66\, mb\,GeV^{-2},\quad A_2 = 4.99\pm 0.57\,
mb\,GeV^{-2}.
\]
The fit result is shown in Fig.~\ref{fig3}. It is seen that the
fitting curve, as in the previous fit \cite{10}, goes excellently
over the experimental points of the CDF group at Fermilab \cite{14}. 

\begin{figure}[htb]
\begin{center}
\begin{picture}(298,184)
\put(20,10){\epsfbox{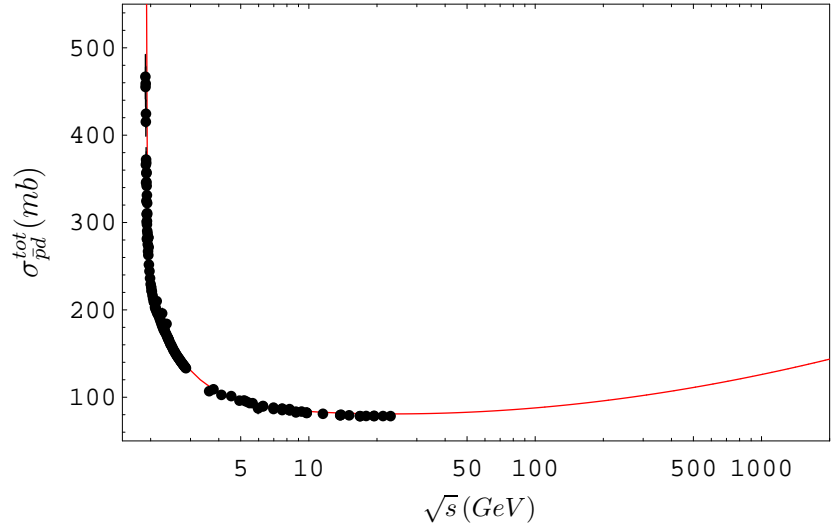}}
\put(155,0){$\sqrt{s}\, (GeV)$}
\put(0,90){\rotate{\large$\sigma^{tot}_{\bar{p}d} (mb)$}}
\end{picture}
\end{center}
\caption{The total antiproton-deuteron cross-section compared with
the theory. Statistical and systematic errors added in
quadrature.}\label{fig4}
\end{figure}

We can substitute $2 \sigma_{sd}^{ex}$ in Eq.~(\ref{30}) for 
formula (\ref{31}), after that the expression for the total shadow
correction may be rewritten in the form
\be
\delta\sigma_{\bar p d}(s) = 
\frac{\sigma_{\bar p N}^{tot\,2}(s)}{4\pi[R^2_d + 2 B_{\bar p
N}^{el}(s)]} +
\frac{A_0 +
A_2\ln^2(\sqrt{s}/\sqrt{s_0})}{R_d^2[1+R_0^2(s)/R_d^2]^{3/2}},
\label{32}
\ee
where all the parameters are fixed according to our previous fits
\cite{10,13} apart from $R_d^2$, which is considered as a single
free fit parameter. Our fit yielded
\[
R_d^2 = 66.61 \pm 1.16 \,GeV^{-2}.
\]
The fit result is shown in Fig.~\ref{fig4}. For completeness the
theory prediction for antiproton-deuteron total cross section is
plotted up to Tevatron energies. 

At this place it should be make the following remark. It is known the
latest experimental value for the deuteron matter radius
$r_{d,m}=1.963(4)\,fm$ \cite{20}. The fitted value for the $R_d^2$
satisfies with a good accuracy to the equality
\be
R_d^2 = \frac{2}{3}r^2_{d,m},\quad  (r^2_{d,m} = 3.853\,fm^2 =
98.96\,GeV^{-2}).
\ee

\begin{figure}[htb]
\begin{center}
\begin{picture}(288,184)
\put(15,10){\epsfbox{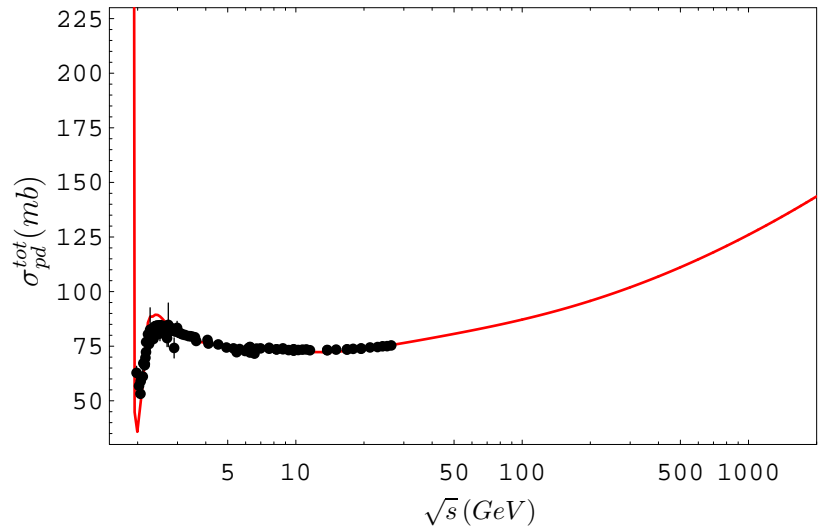}}
\put(155,0){$\sqrt{s}\, (GeV)$}
\put(0,88){\rotate{\large$\sigma^{tot}_{pd} (mb)$}}
\end{picture}
\end{center}
\caption{The total proton-deuteron cross-section compared with the
theory without any free parameters. Statistical and systematic errors
added in quadrature.}\label{fig5}
\end{figure}

Now it would be very intriguing for us to make a comparison of
theoretical formula (\ref{32}) with the data on proton-deuteron total
cross sections where $R_d^2$ has to be fixed by the previous fit to
the data on antiproton-deuteron total cross sections. As in the
previous fit we supposed
$\sigma_{pp}^{tot}=\sigma_{pn}^{tot}\equiv\sigma_{pN}^{tot}$ and
$\sigma_{pp}^{tot}$ had been taken from our global description of
proton-proton total cross sections \cite{10,13}. We also assumed that
$B_{pN}^{el}=B_{\bar p N}^{el}$. So, in this case we have not any
free parameters. The result of the comparison is shown  in
Fig.~\ref{fig5}. As you can see the correspondence of the theory to
the
experimental data is quite remarkable apart from the resonance
region. The resonance region requires a more careful consideration
than that performed here. 

\section{Summary and Discussion}

In this paper we have been concerned with a study of shadow
corrections to the total cross section in scattering from deuteron.
The dynamic apparatus based on the single-time formalism in QFT has
been used  as a tool and subsequently applied to describe the
properties of high-energy particle interaction in scattering from
two-body composite system.  As we have repeatedly emphasized in our
previous works, the conceptual  notion of the new fundamental forces
i.e. three-body forces appeared as a consequence of consistent
consideration of the dynamics for three particle system in the
framework of relativistic quantum theory. In our previous
investigation, we have provided the general framework and described
some general properties of the three-body (in general many-body)
forces to implement the crucial property of any theory such as the
general requirements of unitarity and analyticity \cite{21,22}.
Within this framework we have established a profound relationship of
the three-body forces to the dynamics of one-particle inclusive
reactions. 

The main topic of our studies was to develop the methods which form
the basis for both analytical calculations and phenomenological
investigations. Such developments are necessary for providing an
understanding of the relation between the general structure of the
relativistic quantum theory and relevant hadronic phenomena
described, as a rule, in the frame of the phenomenological models.  

Even though our motivation to construct the general
formalism to study the dynamics of a relativistic three-particle
system has been, in the main, a theoretical one, we have applied this
formalism to investigate the properties of the resultant
hadron-deuteron interaction. 

We have calculated explicitly the contribution of three-body
forces to the total cross section in scattering from any two-body
composite system and investigated the resulting strong interaction
phenomena by applying our approach to the well-known relevant case,
i.e hadron--deuteron scattering. It seems that very weakly bound
two-nucleon state, the deuteron, exhibits the dynamics which leaves
the clustering of the quarks into hadrons essentially intact during
the interaction with the incident hadron and therefore makes, in a
natural way, the dynamical scheme accessible to a description in
terms of nucleonic degrees of freedom only. In this way we found the
new structures for the total shadow correction to the total cross
section in scattering from deuteron. 

First of all, it has been observed that the total shadow correction
inherits the general structure of total cross section and contains
two inherent parts as well, an elastic part and inelastic one. This
partitioning is performed explicitly in the framework of our
approach. It turns out that the elastic part can be expressed
through the elastic scaling (structure) function and the fundamental
dynamical quantity, which is the total elastic cross section in
scattering from an isolated constituent (nucleon) in the composite
system (deuteron). At the same time the inelastic part is expressed
through the inelastic scaling (structure) function and the
fundamental dynamical quantity, which is the total single diffractive
dissociation cross section in scattering from an isolated constituent
in the composite system too. Thus, the general formalism in QFT makes
it possible to define properly the dynamics of particle scattering
from a composite system and express this dynamics in terms of the
fundamental dynamics of particle scattering from an isolated
constituent in the composite system and the structure of the
composite system as itself. We have restricted ourselves to the
simplest composite system, which a two-body composite system
(deuteron) is. However, our general formalism can be
straightforwardly
applied to any multiparticle system and may be used to specify
the dynamics for any many-body composite system as well. There is no,
in principle, difficulty in extending general formalism to more
complex compound many-particle systems such as, for example, nuclei.
We have not attempted to study such extension in this paper, hope,
this will be the subject of our future studies. The main goal of this
work is to gain some insight into hadronic phenomena resulting from
compositeness in the presence of three-body (in general many-body) 
forces.     

The general formalism, which we have outlined, tells us that the
obtained results are substantially more general because they have a 
reliable ground in the framework of the relativistic quantum theory. 
It is evident now that these results correspond to the very deep
physical phenomena in the fundamental dynamics.

What seems most important, which we have discovered in the
work, is that the  elastic and inelastic structure functions have
quite different behaviour. The inelastic structure function has the
maximum and tends to zero at infinity, while the elastic structure
function is the monotonic function and tends to unity at infinity.
This is the most significant difference between the elastic and
inelastic structure functions and it has far reaching physical
consequences. This difference manifests itself in the effect of
weakening of inelastic eclipsing (screening) at super-high energies.
What does it mean physically? To understand it let's combine the
elastic shadow correction and the first term in Eq.~(\ref{28}) for
the hadron-deuteron total cross section 
\be
\sigma_{hd}^{tot} = 2 \sigma_{hN}^{inel} + 2 \sigma_{hN}^{el}(1
-
a^{el}) - \delta\sigma^{inel},\quad
1-a^{el}=\frac{1}{1+x^2_{el}}. \label{34}
\ee
We have in this way that asymptotically
\be
\sigma_{hd}^{tot} = 2 \sigma_{hN}^{inel},\quad s\longrightarrow
\infty.
\ee
Probably the generalization of this result to any many-nucleon
systems (nuclei) looks like
\be
\sigma_{hA}^{tot} = A \sigma_{hN}^{inel},\quad s\longrightarrow
\infty.
\ee
Obviously, this result confirms theoretically the so called quark
counting rules. Moreover, it turns out that the total absorption
(inelastic) cross section manifests itself as a fundamental dynamical
quantity for the constituents in a composite system.

\begin{figure}[htb]
\begin{center}
\begin{picture}(288,188)
\put(20,10){\epsfbox{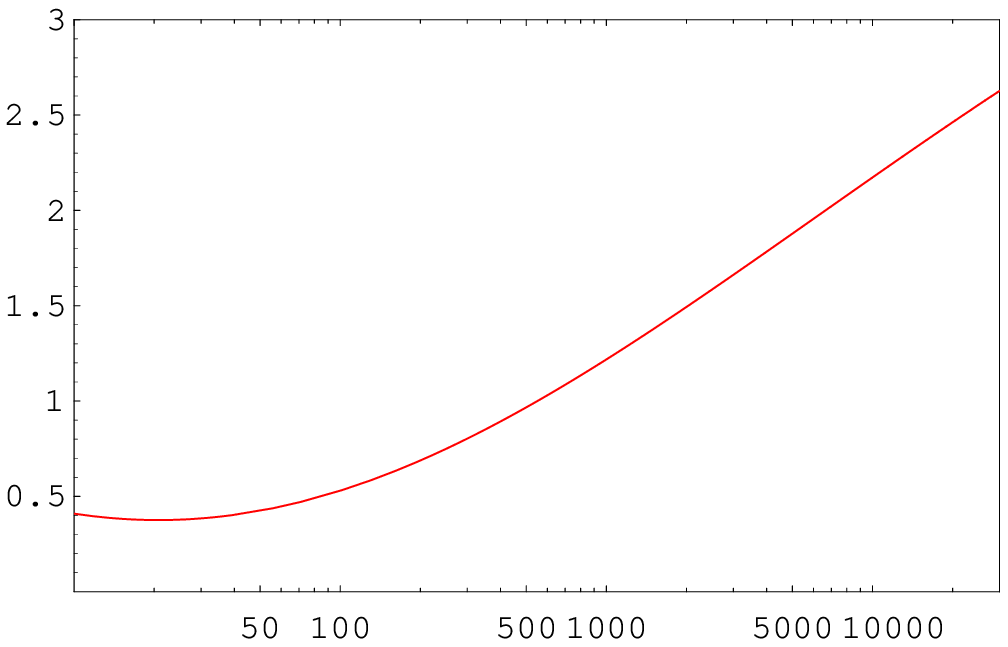}}
\put(155,0){$\sqrt{s}\, (GeV)$}
\put(5,87){\rotate{\large$\delta\sigma^{inel} (mb)$}}
\end{picture}
\end{center}
\caption{The three-body forces contribution (inelastic screening)
to the total antiproton-deuteron cross-section calculated with the
theory.}\label{fig6} 
\end{figure}

\begin{figure}[htb]
\begin{center}
\begin{picture}(288,172)
\put(20,10){\epsfbox{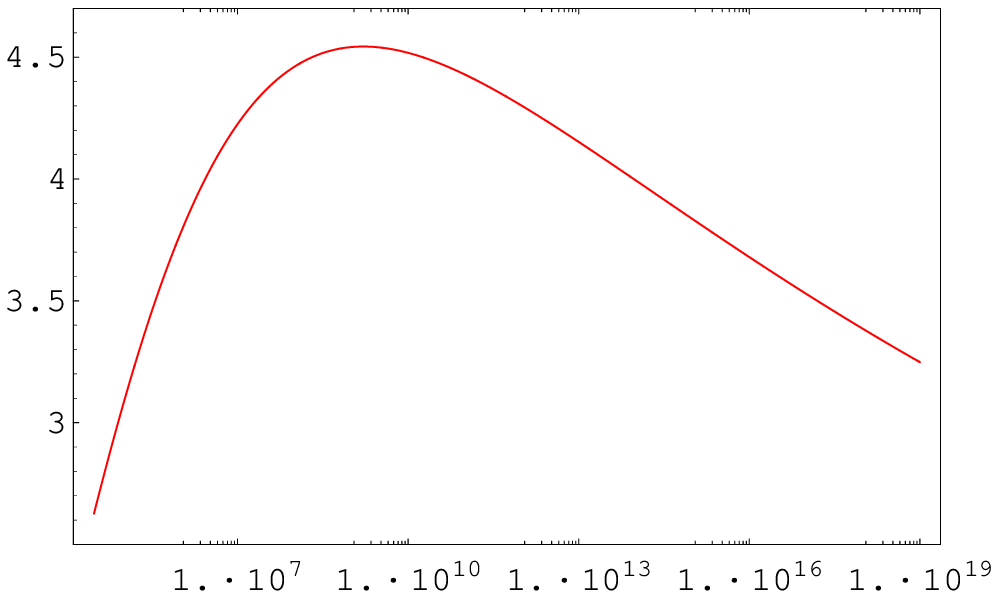}}
\put(155,0){$\sqrt{s}\, (GeV)$}
\put(5,87){\rotate{\large$\delta\sigma^{inel} (mb)$}}
\end{picture}
\end{center}
\caption{The three-body forces contribution (inelastic screening)
to the total antiproton-deuteron cross-section calculated with
the theory in the range up to Planck scale.}\label{fig7} 
\end{figure}

We would also like to emphasize the different range of variation
for the elastic and inelastic structure functions
\be
0 \leq a^{el} \leq 1,\qquad 0 \leq a^{inel} \leq 2/3\sqrt{3}.
\ee
The inelastic shadow correction in a wide range of energies (up to
Planck scale) is shown in Figs.~\ref{fig6},\ref{fig7}. The energy,
where the inelastic shadow correction has a maximum, has to be
calculated from the equality $R_0^2(s_m)=2R_d^2$. Taking
$R_d^2=66.61$ from the fit and $R_0^2(s)$ from paper \cite{13}, we
obtain $\sqrt{s_m}=9.01\,10^8\,GeV=901\,PeV$. Of course, such
energies are not available at now working accelerators. However, we
always have room for a speculative discussion. For example, let us
consider a proton as a two-body (quark-diquark) composite system.
From the experiment it is known that the value for the charge radius
of the proton $r_{p,ch}=0.88\,fm$. If we put $R_p^2=2/3\,r^2_{p,ch}$,
then resolving the equation $R_0^2(s_p)=R_p^2$, we obtain
$\sqrt{s_p}=1681\,GeV$. This is just the energy of Tevatron.
Furthermore in the point $\sqrt{s_0}=20.74\,GeV$ of the minimum for
proton-proton(antiproton) total cross sections, we find
$R_0(s_0)=0.45\,fm$. This is just one half of the proton charge
radius.

So, a realistic and fundamental property of our approach is that it
exhibits two clearly distinct energy regions, associated with the
energies where the range of three-body forces is small compared with
the size of two-body composite system, on the one hand, and with the
energies where the range of three-body forces is large compared with
the two-body bound state size, on the other hand. The size of
two-body compound system plays a role of ``fundamental scale"
separating these two distinct energy regions. 

\begin{figure}[htb]
\begin{center}
\begin{picture}(288,182)
\put(20,10){\epsfbox{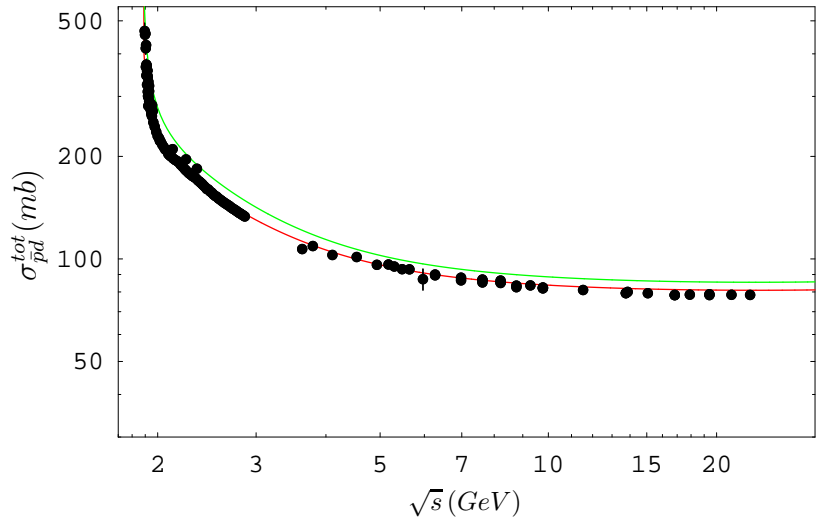}}
\put(155,0){$\sqrt{s}\, (GeV)$}
\put(5,87){\rotate{\large$\sigma^{tot}_{\bar{p}d} (mb)$}}
\end{picture}
\end{center}
\caption{The total antiproton-deuteron cross-section compared with
the theory with and without shadow
corrections.}\label{fig8}
\end{figure}

\begin{figure}[htb]
\begin{center}
\begin{picture}(288,178)
\put(20,10){\epsfbox{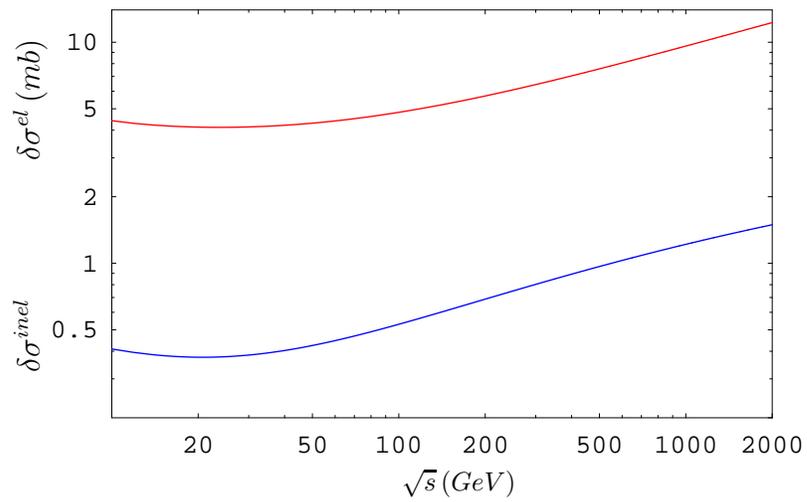}}
\put(155,0){$\sqrt{s}\, (GeV)$}
\put(8,45){\rotate{\large$\delta\sigma^{inel}$}}
\put(8,125){\rotate{\large$\delta\sigma^{el} \,(mb)$}}
\end{picture}
\end{center}
\caption{Elastic and inelastic shadow corrections represented by the
theory.}\label{fig9}
\end{figure}

Fig.~\ref{fig8} displays the significance of shadow corrections in
elastic scattering from deuteron. The elastic and inelastic shadow
corrections to the proton(antiproton)-deuteron total cross sections
are plotted in Fig.~\ref{fig9}. Our analysis shows that the magnitude
of inelastic shadow correction is about 10 percent of elastic one at
available energies. Figuratively speaking, if we called the elastic
shadow corrections a fine structure in the total cross sections, then
we might call the inelastic shadow corrections a super-fine structure
in the total cross sections. In this sense three-body forces make a
``fine tuning" in the dynamics of the relativistic three-body system.
That is why the precise measurements of hadron-deuteron total
cross-sections at high energies are most important. Therefore, it
would be very desirable to think about the creation of accelerating
deuterons beams instead of protons ones at the now working
accelerators and colliders. 

At last, let me remember a unique phenomenon in the history of human
civilization related to Pythagoras, a Greek mathematician and
philosopher, who lived in the sixth century B.C. Gathering together a
group of pupils in the Greek sity of Croton in southern Italy, he
organized a brotherhood devoted to both learning and virtuous living.
The Pythagorean brotherhoods remained active for several centuries. 
The great ideas of Pythagoras and his followers exerted great
influence on the intellectual development of human civilization and
had a fundamental importance at all times. The well known Pythagoras
Theorem, a major step in the devlopment of geometry, is that the
square of the hypotenuse of a right-angle triangle equals the sum of
squares of the two other sides, together with its corollary, namely,
that the diagonal of a square is incommensurable with its side. The
next theorem is that the sum of the angles within any triangle is 180
degrees. Of great influence were the Pythagorean doctrines that
numbers were the basis of all things and possessed a mystic
significance, in particular the idea that the cosmos is a
mathematically ordered whole.  Pythagoras was led to this conception
by his discovery that the notes sounded  by stringed instrument are
related to the length of the strings, he recognized that first four
numbers, whose sum equals $10$ (so called Pythagorean quaternion
$1+2+3+4=10$), contained all basic musical intervals: the octave, the
fifth and the fourth. In fact, all the major consonances, that is,
the octave, the fifth and the fourth are produced by vibrating
strings whose lengths stand to one another in the ratios of $1:2$,
$2:3$ and $3:4$ respectively. The resemblance which he perceived
between the orderlines of music, as expressed in the ratios which he
had discovered and the idea that cosmos is an orderly whole, made up
of parts harmoniously related to one another, led him to conceive of
the cosmos too as mathematically ordered. The Pythagoreans supposed
that the universe was a sphere in which the planets revolved. The
revolving planets were thought to produce musical notes -- ``the
music of the spheres". The importance of this conception is very
great, for example, it is the ultimate source of Galileo's belief
that ``the book of nature is written in mathematical symbols" and
hence the ultimate source of modern physics in the form in which it
came to us from Galileo. The Pythagoreans believed also in
reincarnation, that is, the soul, after death, passes into another
living thing, which presupposes the ability of the soul to survive
the death of the body, and hence some sort of belief in its
immortality.

As it was established above in our study the inelastic structure
function $a^{inel}$ has the maximum and at the maximum this function
equals $2/3\sqrt{3}$. The number $2/3\sqrt{3}$ may be considered as a
fundamental number calculated in the theory with a clear physical
interpretation. We also found the relations $R^2_0(s_m):R^2_d=2:1$
and $R_0(s_0):r_{p,ch}=1:2$ which looked like harmonic ratios
mentioned above and hence might be considered as ``the music"
produced by diffraction phenomena in high energy elementary particle
physics. It seems, we come back to the great Pythagorean ideas
reformulated in terms of the objects living in the microcosmos. The
great Pythagorean idea applied to the microcosmos might be shown by
the following diagram:
\vskip 0.1 true in
\begin{center}
{\Large
$\framebox{\bf DIFFRACTION PHENOMENA}$\\
\vskip 0.1 true in
$\Downarrow$
\vskip 0.1 true in
$\framebox{\bf THE MUSIC (HARMONY) OF THE SPHERES}$
\vskip 0.1 true in
$\Updownarrow$
\vskip 0.1 true in
$\framebox{\bf THE HARMONY (MUSIC) OF THE NUMBRERS}$}
\end{center}
\vskip 0.1 true in
So, it appears that the study of diffraction phenomena in high energy
elementary particle physics makes it possible to establish a missing
link between cosmos and microcosmos, between the great ancient ideas
and recent investigations in particle and nuclear physics and to
confirm the unity of physical picture of the World. Anyway, we
believe in it.

\section*{Acknowledgements}
It is my great pleasure to express thanks to the Organizing 
Committee for the kind invitation to attend this wonderful Workshop.
I would like to especially thank V.I. Savrin, V. Keshek and all local
organizers for warm and kind hospitality throughout the Workshop.

\newpage
\vspace*{2cm}
{\Large
\[
\frac{s}{\pi}\frac{d\sigma_{hN\rightarrow NX}}{dtdM_X^2} =
\frac{(2\pi)^3}{I(s)}\chi(\bar s)Im{\cal F}_0(\bar
s;-\vec{\Delta}, \vec{\Delta}, \vec q; 
\vec{\Delta}, -\vec{\Delta}, \vec q\,)
\]
\[
\bar s = 2(s + M_N^2) - M_X^2,\quad t = - 4{\vec\Delta}^2.
\]
\centerline{($I(s)/\chi(\bar s)$ -- ``renormalized flux"!)}
}

\newpage
\vspace*{2cm}
\vspace{1cm}

\newpage
\begin{table}
\caption{Data on $p\bar p$ single diffraction dissociation
cross-sections.}\label{tab}
\begin{center}
\begin{tabular}{|c|c|c|}\hline
$\sqrt{s}\ (GeV)$ & $\sigma^{p\bar p}_{sd}(mb)$ & References \\
\hline
    14.00   & $ 3.94  \pm 0.20   $ & \cite{15} \\ \hline
    16.20   & $ 4.87  \pm 0.08   $ & \cite{15} \\ \hline
	17.60   & $ 4.96  \pm 0.08   $ & \cite{15} \\ \hline
    19.10   & $ 4.94  \pm 0.08   $ & \cite{15} \\ \hline
    20.00   & $ 4.46  \pm 0.25   $ & \cite{15} \\ \hline
    20.00   & $ 4.9   \pm 0.55   $ & \cite{15} \\ \hline
    23.30   & $ 6.50  \pm 0.2    $ & \cite{15} \\ \hline
    23.40   & $ 6.07  \pm 0.17   $ & \cite{15} \\ \hline
    23.80   & $ 5.19  \pm 0.08   $ & \cite{15} \\ \hline
    26.90   & $ 6.05  \pm 0.22   $ & \cite{15} \\ \hline
    27.20   & $ 5.42  \pm 0.09   $ & \cite{15} \\ \hline
    27.40   & $ 6.30  \pm 0.2    $ & \cite{15} \\ \hline 
    30.50   & $ 6.37  \pm 0.15   $ & \cite{15} \\ \hline
    32.30   & $ 6.32  \pm 0.22   $ & \cite{15} \\ \hline
    32.40   & $ 6.50  \pm 0.2    $ & \cite{15} \\ \hline
    35.20   & $ 7.01  \pm 0.28   $ & \cite{15} \\ \hline
    35.50   & $ 7.50  \pm 0.5    $ & \cite{15} \\ \hline
    38.30   & $ 6.08  \pm 0.29   $ & \cite{15} \\ \hline
    38.50   & $ 7.30  \pm 0.4    $ & \cite{15} \\ \hline
    44.70   & $ 7.30  \pm 0.3    $ & \cite{15} \\ \hline
    53.70   & $ 7.00  \pm 0.3    $ & \cite{15} \\ \hline
    62.30   & $ 7.50  \pm 0.3    $ & \cite{15} \\ \hline
    200     & $ 4.8  \pm  0.9    $ & \cite{16} \\ \hline
    546     & $ 5.4  \pm  1.1    $ & \cite{17} \\ \hline
    546     & $ 7.89  \pm  0.33  $ & \cite{14}  \\ \hline
    546     & $ 9.4  \pm  0.7    $ & \cite{18} \\ \hline
    546     & $ 8.34 \pm 0.36    $ & \cite{15} \\ \hline
    900     & $ 7.8   \pm  1.2   $ & \cite{18} \\ \hline
    1800    & $ 9.46  \pm  0.44  $ & \cite{14} \\ \hline
    1800    & $ 11.7  \pm  2.3   $ & \cite{19} \\ \hline
    1800    & $ 8.1  \pm  1.7    $ & \cite{19} \\ \hline
    1800    & $ 8.46 \pm 1.77    $ & \cite{15} \\ \hline
    1800    & $ 9.12 \pm 0.46    $ & \cite{15} \\ \hline
\end{tabular}
\end{center}
\end{table}

\newpage
\vspace*{2cm}
\vspace{1cm}
\newpage
\vspace*{2cm}
\vspace{1cm}
\newpage
\vspace*{2cm}
\vspace*{1cm}
\end{document}